\documentclass{aastex}

\usepackage{psfig,amsmath,emulateapj5}

\shorttitle{Quirrenbach et al.}
\shortauthors{Keck Adaptive Optics Observations of 3C294}

\begin{document}


\title{Keck Adaptive Optics Observations of the Radio Galaxy 3C294: \\
A Merging System at $z = 1.786$?}


\author{A. Quirrenbach\altaffilmark{1}, J.E. Roberts, K.
Fidkowski\altaffilmark{2}}
\affil{University of California, San Diego, Center for Astrophysics and Space
Sciences, La Jolla, CA 92093-0424}
\email{aquirrenbach@ucsd.edu}

\and

\author{W. de Vries\altaffilmark{1}, W. van Breugel\altaffilmark{1}}
\affil{Institute of Geophysics and Planetary Physics, Lawrence Livermore
National Laboratory, Livermore, CA 94550-9900}


\altaffiltext{1}{Visiting Astronomer, W.M. Keck Observatory}
\altaffiltext{2}{REU Summer Student}


\begin{abstract}
We present seeing-limited and adaptive optics (AO) images of the $z = 1.786$
radio galaxy 3C294 in the $H$ and $K'$ infrared bands obtained at Keck
Observatory. The infrared emission of 3C294 is dominated by two distinct
components separated by $\sim 1''$ (9\,kpc). The eastern knot contains an
unresolved core that contributes $\sim 4\%$ of the $K'$-band light; we identify
this core with the active nucleus. The western component is about 2.5 times
brighter. The most plausible interpretation of the near-infrared morphology is
an ongoing merger event, with the active nucleus located in the less massive of
the two galaxies.
\end{abstract}



\section{Introduction}

Powerful radio galaxies offer a unique way to investigate the evolution of very
massive galaxies over a large range in redshift. The near-infrared Hubble $K -
z$ relation is well represented by models of passive evolution of massive ($5 -
10\,L^\ast$) galaxies which formed at high redshifts (Lilly 1988, van Breugel
et al.\ 1999). Comparison with high-redshift field galaxies (Cowie 1997)
confirms that radio galaxies form indeed the high-luminosity envelope. Deep
spectroscopic observations of a few relatively weak radio sources, in which the
active nuclei do not dominate the restframe UV radiation, have shown directly
that radio galaxies at $z \approx 1.5$ have old ($\gtrsim 3.5 - 4.5$\,Gyr)
stellar populations, which must have formed at $z \gtrsim 10$ (Spinrad et al.\
1997).

Radio galaxy morphologies, when imaged at visible wavelengths, often show
spectacular clumpy structures aligned with the axis of the radio source. This
``alignment effect'' appears to be at odds with the tightness of the
near-infrared $K - z$ relation and the passive evolution inferred from the $K -
z$ diagram. Its exact nature has remained unclear and evidence has been found
for scattered light from hidden quasar-like active nuclei, for nebular
recombination continuum, and even for jet-induced star formation (McCarthy
1993).

To investigate the morphological evolution of massive elliptical galaxies with
redshift it is therefore important to obtain high spatial resolution at
infrared wavelengths, where the AGN-related emission is fainter and the old
stellar component brighter. Recent $H$-band observations with NICMOS/HST have
shown that at $z < 2$ most galaxy morphologies appear rounder, more symmetric
and centrally concentrated than at shorter wavelengths, although even at
restframe R-band wavelengths some of the complex aligned structures can still
be seen (Zirm et al.\ 1999). After subtracting these aligned structures and
modeling the symmetric galaxy components, the galaxy surface brightness
profiles show $r^{1/4}$ power laws, consistent with the relaxed morphological
appearances. In several cases the images also show that the near-IR surface
brightness peaks at local minima in WFPC2 optical images, suggesting the
presence of central dust lanes or disks. Some objects exhibit relatively bright
nuclear point sources in the near-IR, possibly showing obscured quasar-like
AGN.

NICMOS/HST observations of radio galaxies at $z > 2$ generally reveal much more
complex structures than for their lower redshift counterparts (Pentericci et
al.\ 1999, McCarthy 1999). These sources have more often strong nuclear point
sources, show nearly always complex and asymmetric morphologies, and exhibit
the alignment effect. This is in part because the observations of these higher
redshift objects are necessarily at shorter restframe wavelengths, but also
because of strong morphological evolution. $K$-band images of high-redshift
radio galaxies obtained with the NIRC camera at Keck Observatory in $0 \farcs
4$ to $0 \farcs 7$ seeing have also shown dramatic evidence for this, with
multiple $L^\ast$ components which are often aligned with their radio sources
at $z > 3$, to more symmetric and compact morphologies at $z < 3$ (van Breugel
et al.\ 1998). Such morphological evolution, from clumpy (but not necessarily
aligned) to more symmetric and relaxed structures, would be expected in
hierarchical galaxy formation scenarios. It is thus clear that near-IR
observations at high resolution can substantially contribute to our
understanding of the structures, evolution, and environments of radio galaxies.

3C294 is a powerful radio galaxy at $z=1.786$; a VLA image at $\lambda = 6$\,cm
shows a Z-shaped structure with a relatively weak core (McCarthy et al.\ 1990).
The galaxy is associated with a large cloud of ionized gas radiating in
Lyman~$\alpha$ and other emission lines. The cloud is aligned with the inner
part of the radio structure, and extends over $\sim 12''$ (McCarthy et al.\
1990). At the redshift of 3C294, 1$''$ corresponds to 9.1\,kpc for $H_0 =
65$\,km\,s$^{-1}$\,Mpc$^{-1}$, $\Omega_0 = 0.3$, $\Omega_\Lambda
= 0.7$; the linear extent of the Lyman~$\alpha$ cloud is thus $\sim 110$\,kpc.

A bright ($V = 12$) star is projected $10''$ west of the radio galaxy core.
3C294 is thus an ideal target for observations with present-day adaptive optics
(AO) systems, which require a bright natural reference star near the target
source. Previous AO observations of 3C294 have been carried out in the $K'$
band with the Hokupa'a system mounted on the 3.6\,m Canada-France-Hawaii
Telescope (Stockton et al.\ 1999). The galaxy was clearly resolved in these
observations, but the core could not be identified unambiguously. The overall
morphology appeared triangular, and was interpreted by Stockton et al.\ (1999)
as an illumination cone due to dust scattering from a quasar nucleus.

Here we present new seeing-limited and AO observations of 3C294 in the $H$ and
$K'$ near-IR bands obtained at Keck Observatory. The greater sensitivity and
resolution afforded by a 10\,m aperture gives substantially improved
morphological information, which results in an improved understanding of the
origin of the emission, and of the evolutionary state of 3C294.

\section{Observations and Data Reduction}

\subsection{NIRC Observations}

Direct infrared images of 3C294 were obtained on June 25, 2000 (UT date) with
the NIRC camera on the Keck\,I telescope. NIRC's detector is a $256 \times 256$
pixel InSb array; the field-of-view is $38 \farcs 4 \times 38 \farcs 4$ with a
sampling of $0 \farcs 15$ per pixel (Matthews \& Soifer 1994). We used an
exposure time of 10\,s to avoid saturation by the bright star $10''$ west of
3C294, and observed at a parallactic angle at which the radio galaxy is well
separated from the diffraction spikes caused by the support structure of the
secondary mirror of the telescope (see Figure~\ref{NIRCK}). We collected 54
frames each in the infrared $H$ and $K'$ bands, giving total on-source
integration times of 9 minutes in each band. The source was dithered on the
detector in a nine-point pattern to allow for easy sky correction and
elimination of bad pixels. The UKIRT faint standard star FS23 was observed
immediately before 3C294. The seeing was fairly good ($\sim 0 \farcs 6$ to
$\sim 0 \farcs 7$), but the observations were affected by cirrus and are not
photometric.

Standard infrared data reduction techniques were applied to the data. The sky
emission in each pixel was determined from the median of the nine images in
each dither pattern and subtracted before the nine images were centered and
co-added. The resulting mosaic images were added and divided by a flat field
obtained on the sky during twilight.

\subsection{Adaptive Optics Observations}

High-resolution infrared observations of 3C294 were carried out on June 24,
2000 (UT date). We used the adaptive optics system mounted on the Keck\,II
telescope, which uses a Shack-Hartmann sensor and a fast low-noise CCD camera
for wavefront sensing, and a Xinetics mirror with 349 actuators for wavefront
correction (see Wizinowich et al.\ 2000). The $V=12$ star $10''$ west of 3C294
was used as wavefront reference for the AO system. Infrared images were
obtained with the slit viewing camera (SCAM) of the NIRSPEC spectrograph
(McLean et al.\ 2000). SCAM employs a $256 \times 256$ pixel {\sc Picnic}
HgCdTe array; the pixel scale for observations in conjunction with the AO
system is $0 \farcs 017$/pixel, giving a field-of-view of $4 \farcs 3 \times 4
\farcs 3$. Since this field is too small to observe 3C294 and the guide star
simultaneously, we periodically moved back to the guide star to assess the AO
system performance by measuring its Strehl ratio.

The exposure times were 300\,s for 3C294, and 2\,s for the guide star; a
five-point dither pattern with a step size of $0 \farcs 75$ was used in both
cases. The dithering and switching between 3C294 and the guide star was done by
moving the telescope on the sky, and performing compensating changes in the
position of a motorized field steering mirror located behind the visible/IR
beam splitter in the wavefront sensor arm of the optics. We obtained 20
individual images of 3C294 in $H$ and 15 images in $K'$, i.e., the total
on-source integration times were 100 and 75 minutes, respectively. The UKIRT
faint standard star FS29 was observed after 3C294. The seeing was good ($\sim 0
\farcs 5$) and the sky clear. Under these conditions, the AO system produced
images with diffraction-limited cores ($\sim 0 \farcs 047$) and a Strehl ratio
of $\sim 30\%$ in $H$ on $V=12$ reference stars.

The data reduction was done essentially in the same way as for the NIRC data. A
slight complication arose from the fact that we were using the slit viewing
camera of a spectrograph. Since no closed position of the slit mechanism was
provided, we used the narrowest slit available ($0 \farcs 013 \times 1 \farcs
13$), and masked the corresponding area in the center of the chip in the data
reduction. This was not a major problem because of the dither pattern, which
ensured that each sky position fell onto a valid area of the chip in at least
four out of five exposures.

Since the bright star was not in the field-of-view of the observations of
3C294, and the radio galaxy was not detected with sufficient signal-to-noise in
the individual exposures, the registration of the 20 frames had to be done
``blindly''. The AO system keeps the guide star at a fixed position on the
wavefront sensor camera, and the offset of the SCAM detector with respect to
the wavefront sensor can be derived from the information about the position of
the field steering mirror in the FITS headers. A correction has to be applied
for differential refraction between the effective wavelength of the wavefront
sensor camera ($\sim 750$\,nm) and the $H$ band.

\section{Results}

\begin{figure*}
\caption{\label{NIRCK}
Keck NIRC $K'$-band image of 3C294. The bright star $\sim 10''$ west of the
radio galaxy was used as guide star for the adaptive optics
observations.\vspace{5mm}}
\psfig{figure=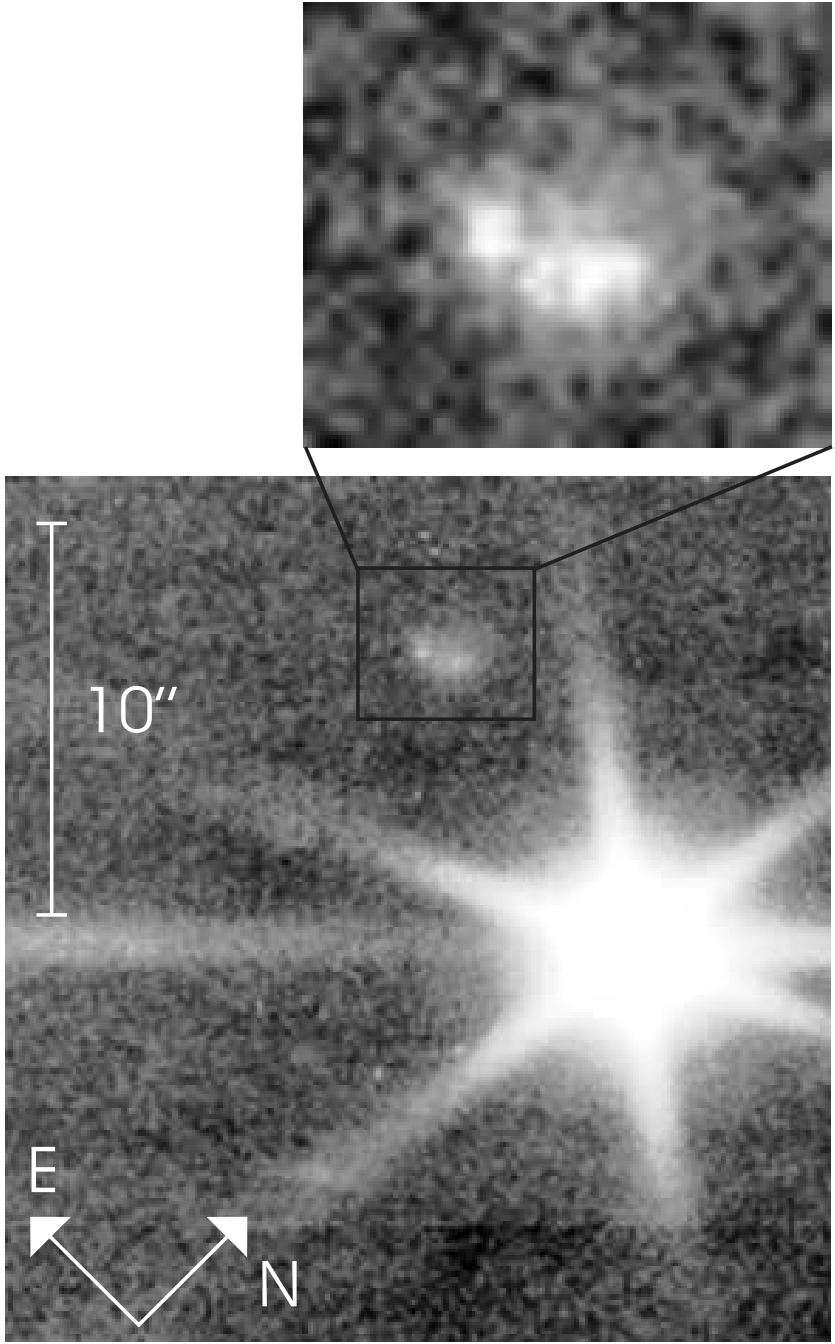,width=11cm}
\end{figure*}

\begin{figure*}
\caption{\label{AOH}
$H$-band image of 3C294 obtained with the slit-viewing camera (SCAM) of NIRSPEC
behind the Keck adaptive optics system. North is up and East to the left. The
bright core of the eastern component is unresolved; its diameter is $\leq 0
\farcs 05$.\vspace{5mm}}
\psfig{figure=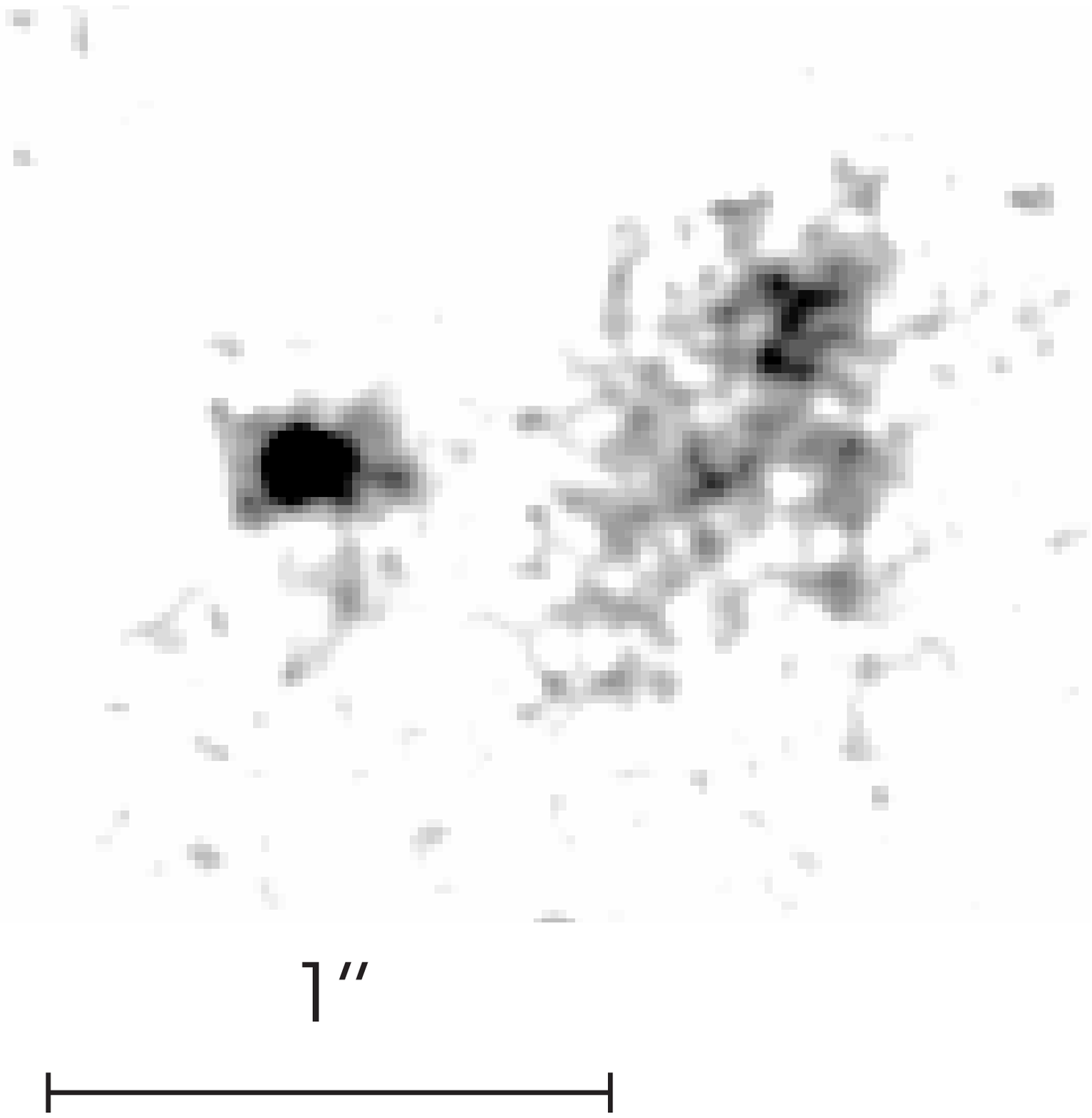,width=11cm}
\end{figure*}

The final NIRC $K'$-band and AO $H$-band images of 3C294 are shown in
Fig.~\ref{NIRCK} and \ref{AOH}, respectively. Note that no deconvolution has
been applied in either case. Two bright components separated by $\sim 1''$ are
apparent in both images; the NIRC image also shows faint diffuse emission. The
AO image shows that the eastern component (the ``core'') is much more compact
than the western component. The NIRC $H$-band and AO $K'$-band images are very
similar to the images shown, but have slightly poorer signal-to-noise.

Adopting $H = 12.4$, $K' = 12.3$ for FS23 and $H = 13.3$, $K' = 13.3$ for FS29,
we derive $H = 19.4$, $K' = 18.2$ for the integrated emission of 3C294, and $H
= 22.0$, $K' = 21.7$ for the core. We estimate photometric errors of at least
$\pm 0.2$\,mag due to the non-photometric conditions, but our measurements are
in excellent agreement with previous results ($K = 18.0 \pm 0.3$, McCarthy et
al.\ 1990; $K' = 18.3 \pm 0.3$, Stockton et al.\ 1999). A comparison of the
photometric data from NIRC and AO gives the interesting result that only $\sim
9$\% of the $H$ flux and $\sim 4$\% of the $K'$ emission come from the
``core''.

The width of the core can be determined by comparing the AO image of 3C294 to
the exposures of the guide star. The guide star is a double star with 1:1.6
intensity ratio and $0 \farcs 15$ separation (at the time of our observations).
It is thus well resolved in our AO data, and it is easy to obtain radial
profiles for both components separately, i.e., the duplicity is not a problem.
It is necessary, however, to check the quality of the registration of the 20
individual exposures that were stacked to obtain Fig.~\ref{AOH}. We performed
this test by building similar mosaics composed of exposures of the guide stars.
We first stacked guide star exposures in the same way as we did for 3C294,
i.e., by using the information about the position of the field steering mirror
in the FITS headers to register the individual frames. The resulting mosaic
image had a FWHM of 2.78 pixels, corresponding to $0 \farcs 0473$. We then
stacked the same frames by computing cross-correlations between the individual
exposures; the mosaic constructed in this way had a FWHM of 2.79 pixels or $0
\farcs 0474$, identical to within the errors to the first. We conclude that
``blind'' stacking based on the field steering mirror positions did not have a
measurable effect on the FWHM of the resulting mosaic.

The FWHM of the eastern component of 3C294 was measured to be 3.38 pixels,
i.e., $0 \farcs0575$. Quadratic subtraction of the FWHM of the guide star gives
a formal size of $0 \farcs 033$. We also compared radial profiles of the
eastern component with profiles of the guide star that had been convolved with
Gaussians of varying widths. From these comparisons we derive a conservative
upper limit of $0 \farcs 05$ for the FWHM of the core of 3C294. It should be
noted that these measurements are still consistent with the core being
unresolved, since angular anisoplanatism, the fairly low signal-to-noise ratio
in Fig.~\ref{AOH}, or an underlying emission from a larger stellar component
could all contribute to an apparent broadening of the core compared to the
guide star.

An important question concerns the registration of the infrared emission with
respect to the radio structure. We can measure the position of the infrared
core (which can clearly be identified in Fig.~\ref{NIRCK}) with respect to the
guide star from the NIRC $K'$ data; it is $9 \farcs 5$ east and $1 \farcs 5$
north of the star, with an estimated error of about $0 \farcs 2$. Performing
the same measurement with the AO data is more complicated, since we have to
rely on the calibration of the field steering mirror (which we checked and
found to be better than 1\%), and because the two components of the guide star
are resolved. The offset between the $H$-band core and the center-of-light of
the two components of the guide star determined from the AO observations is $9
\farcs 7$ east and $1 \farcs 6$ north, in excellent agreement with the NIRC
value. In the following we will use the result of the more straightforward NIRC
measurement.

The guide star is star U1200-07227692 in the USNO-A2.0 catalog (Monet et al.\
1998). Several attempts have been made to determine the position of this star,
with results disagreeing among each other by $\pm 1''$ (V\'eron 1966, Kristian
et al.\ 1974, Riley et al.\ 1980, Stockton et al.\ 1999). We believe that the
best estimate is the USNO-A2.0 catalog position 14$^{\rm h}$06$^{\rm
m}$43$\fs$32, +34$^\circ$11$'$23$\farcs$5 (J2000 on the ICRS reference frame),
derived directly from a solution of the Schmidt plates on which USNO-A2.0 is
based. The typical error of the catalog is $0 \farcs 25$. Combining this
position with the offset determined above, the position of the infrared core of
3C294 is found to be 14$^{\rm h}$06$^{\rm m}$44$\fs$09,
+34$^\circ$11$'$25$\farcs$0. The position of the radio core in the ICRS frame
is 14$^{\rm h}$06$^{\rm m}$44$\fs$08, +34$^\circ$11$'$25$\farcs$0 (McCarthy et
al.\ 1990, Stockton et al.\ 1999). The almost perfect agreement between the
positions of the infrared and radio cores may be somewhat fortuitous in view of
the slightly discrepant positions of the guide star in the literature.
Nevertheless, the positional agreement, as well as the pointlike appearance of
the infrared core, strongly suggest the identification of this component with
the active nucleus.

\section{Discussion}

Our seeing-limited and adaptive optics images of 3C294 are the first to show
sufficient detail for a comprehensive interpretation of the nature of the
near-infrared emission of this galaxy. Most notably, we detect two clearly
distinct knots separated by $\sim 1''$ (9\,kpc) in the east-west direction,
i.e., nearly perpendicular to the radio axis. The eastern knot contains an
unresolved (or barely resolved) core, which we identify with the active
nucleus.

Our AO image is in general agreement with the one obtained by Stockton et al.\
(1999) from their Hokupa'a AO observations. However, because of our higher
quality data, we can now clearly identify the nucleus, which appears offset by
$\sim 1''$ to the east from the main body of the galaxy. This leads us to a
different conclusion from Stockton et al.\ (1999), who suggested that the
morphology of 3C294 might be due to scattered light from a hidden quasar-like
nucleus, located near the southern tip of the $K$-band emission. From our new
images it appears much more plausible to ascribe the bulk of the near-IR
emission to an old stellar population, with a $\sim 4\%$ contribution by the
active nucleus.

The near-IR color $H-K' = 1.2$ of 3C294 is typical for galaxies at redshift
$\sim 1.8$. The $K$-band K-correction $K_K(z)$ is relatively small and does not
depend strongly on redshift or type of galaxy (e.g., Cowie et al.\ 1994);
assuming $K_K (1.8) = -0.5$ we obtain $M_K = -27.1$ for 3C294. Comparing this
value to $M_{K^\ast} = -25.1$ (Mobasher et al.\ 1993), we derive $L = 6.3\,
L^\ast \approx 1.3 \times 10^{11} L_\odot$. With the ``standard''
mass-to-light-ratio $\mathcal{M}/L = 15$, the total mass of 3C294 is then
estimated to be $\mathcal{M} \approx 2 \times 10^{12} \mathcal{M}_\odot$.
McCarthy et al.\ (1990) have used spatially resolved observations of the
Lyman~$\alpha$ line to derive a dynamical estimate of $\mathcal{M} \approx 3
\times 10^{12} \mathcal{M}_\odot$ enclosed within 60\,kpc, but newer data
(McCarthy et al.\ 1996) bring this estimate up to $\mathcal{M} \approx 2.4
\times 10^{13} \mathcal{M}_\odot$ within 90\,kpc, which would imply
$\mathcal{M}/L \approx 200$. These dynamical mass estimate rest on somewhat
shaky ground, however, since it is not clear at all that the emission line
kinematics in 3CR galaxies reflect the underlying gravitational fields (Baum \&
McCarthy 2000), and the distorted structure apparent in our near-IR images of
3C294 may cast further doubts on the applicability of simple virial arguments.

The near-infrared morphology of 3C294 revealed by our Keck images is indicative
of an ongoing merger event, consistent with the expectation from hierarchical
models of structure formation. These models predict that many present-day
massive galaxies have merged with a galaxy of nearly equal mass at $z \leq 2$
(Kauffmann \& White 1993). It is tempting to speculate that the galaxy-galaxy
interaction may have triggered the radio activity, and that it may be
responsible for the distortion (Z-shape) of the radio structure. Assuming an
advance speed of the radio hot spots of $\sim 10,000$\,km\,s$^{-1}$ (McCarthy
et al.\ 1990), the age of the radio source is of order $10^7$\,yr. This is to
be compared with the dynamical time scale of the merger, which is $\sim
10^8$\,yr. The order-of-magnitude difference in the two time scales may
indicate that it takes some time for the gas to settle around the black hole,
after which the accretion proceeds on a shorter time scale. The radio and
infrared morphologies of 3C294 appear therefore consistent with a scenario in
which supermassive black holes grow through major merger events, in which the
gas is accreted on a relatively short time scale, about $10^7$\,yr (Kauffmann
\& Haehnelt 2000).

It is interesting to note that the active nucleus seems to be associated with
the less luminous galaxy. Quantitative photometry is difficult because the AO
data have insufficient surface brightness sensitivity to fully capture the
extended emission, and the NIRC data insufficient resolution to fully separate
the two components. Our best estimate is a $\sim 2.5:1$ ratio for the stellar
$K'$-band luminosities of the western and eastern components.

Our observations clearly show that the $K'$-band luminosity of 3C294 is
dominated by stars, not by the active nucleus. In fact the 4\% of the $K'$-band
light contained in the compact ``core'' sets a stringent upper limit to the AGN
contribution. This agrees with results from lower-resolution imaging of 3CR
galaxies, which indicate that few have nuclei contributing more than $\sim
15\%$ of the $K$-band light (Best et al.\ 1997). It is the dominance of the old
stellar population that produces a well-defined $K - z$ relation, and the locus
of 3C294 in the $K - z$ diagram is indeed close to that of other 3CR galaxies
at comparable redshifts (Best et al.\ 1998). On the other hand, this is
surprising in view of the peculiar near-IR morphology of 3C294. If the activity
of radio galaxies is generally tied to galaxy-galaxy interaction, one should
not expect to find a uniform population of hosts in radio-selected samples, and
therefore a large scatter in the $K - z$ relation. The fact that the radio
source appears to be associated with the less massive of the two galaxies
further complicates the picture, and appears to contradict models based on
passive evolution at $z \lesssim 2$.

Near-IR imaging of a larger sample of radio galaxies with high angular
resolution is clearly needed to clarify these issues. Adaptive optics with
laser guide stars on 10\,m class telescopes will be the ideal tool to perform
these observations.

\acknowledgments

Many thanks are due to the adaptive optics group, instrument specialists, and
observing assistants at W.M.\ Keck Observatory for their excellent advice and
assistance. This work has been supported in part by the National Science
Foundation Science and Technology Center for Adaptive Optics, managed by the
University of California at Santa Cruz under cooperative agreement No.\
AST-98-76783. The work by WdV and WvB was performed under the auspices of the
U.S.\ Department of Energy by University of California Lawrence Livermore
National Laboratory under contract No.\ W-7405-Eng-48. KF's work was supported
by the NSF through REU grant PHY9732035 to the UCSD Physics Department. The
data presented herein were obtained at the W.M.\ Keck Observatory, which is
operated as a scientific partnership among the California Institute of
Technology, the University of California and the National Aeronautics and Space
Administration. The Observatory was made possible by the generous financial
support of the W.M.\ Keck Foundation.





\clearpage

\end{document}